\documentclass[10pt]{wlscirep}
\usepackage[utf8]{inputenc}
\usepackage[T1]{fontenc}
\usepackage{amsthm}
\usepackage{amsmath}
\usepackage{array}
\title{Temporal Link Prediction via Adjusted Sigmoid Function and 2-Simplex Sructure}

\author[1]{Ruizhi Zhang}
\author[2]{Qiaozi Wang}
\author[1]{Qiming Yang}
\author[1,3,*]{Wei Wei}
\affil[1]{LMIB and School of Mathematical Sciences, Beihang University, Beijing, 100191, China}

\affil[2]{Longyuan (Beijing) Wind Power Engineering Technology Co., LTD., Beijing, 100081, China}
\affil[3]{Peng Cheng Laboratory, Shenzhen, Guangdong, 518000, China}

\affil[*]{weiw@buaa.edu.cn}

\begin{abstract}
Temporal network link prediction is an important task in the field of network science, and has a wide range of applications in practical scenarios. Revealing the evolutionary mechanism of the network is essential for link prediction, and how to effectively utilize the historical information for temporal links and efficiently extract the high-order patterns of network structure remains a vital challenge. To address these issues, in this paper, we propose a novel \underline{t}emporal \underline{l}ink \underline{p}rediction model with adjusted \underline{s}igmoid function and 2-\underline{s}implex structure (TLPSS). The adjusted sigmoid decay mode takes the active, decay and stable states of edges into account, which properly fits the life cycle of information. Moreover, the latent matrix sequence is introduced, which is composed of simplex high-order structure, to enhance the performance of link prediction method since it is highly feasible in sparse network. Combining the life cycle of information and simplex high-order structure, the overall performance of TLPSS is achieved by satisfying the consistency of temporal and structural information in dynamic networks. Experimental results on six real-world datasets demonstrate the effectiveness of TLPSS, and our proposed model improves the performance of link prediction by an average of 15\% compared to other baseline methods.
\end{abstract}
\begin{document}

\flushbottom
\maketitle

\thispagestyle{empty}

\section*{Introduction}

With the rapid development of internet and network science, the explosive growth of various information volumes in different fields is accompanied by an urgent need for scientific research and industry to improve data processing capabilities. The complex network, which takes big data and complex associations among data as the research object, is the basic algorithmic framework for modeling data\cite{complex_network, big_data}. Link prediction in complex networks has been widely regarded as one of the most interesting problems in the information field\cite{lp_df}.

The mission of link prediction is to predict the connection  possibility of nodes that have not yet connected in the network, including the recovery of missing links and the formation of future links. The major difference of the aforementioned tasks is that the latter one mainly focuses on dynamic networks, which means links in those networks emerge at different time. 
For example, for protein networks as static data\cite{pp_net}, due to the insufficiency of our empirical knowledge, prediction of two proteins' interaction can be thought as restoration of missing links. Static link prediction focuses on the completeness of the graph, while dynamic link prediction mainly predicts the formation of future links in order to simulate network evolution. It is well established that networks are highly dynamic objects with inherent dynamic properties\cite{dynamic_properties}. Temporal link prediction aims to capture those properties in dynamic networks. It intends to extract the implicit driving force in the network and achieve the goal of network evolution analysis\cite{dynamic_lp}. The most important application of it is in recommender systems\cite{recommender}, which have been widely used in many fields, such as e-commerce, social network and other scenarios\cite{electronic_commerce, social_network}. 

Among all successful link prediction methods, similarity method is one of the most commonly used link prediction methodology. However, traditional similarity methods solely consider the current static state of networks, such as the topology structure, while ignoring the temporal dimensional evolution pattern of complex networks\cite{similarity}. This type of method is not suitable for temporal networks, where edges are annotated with timestamps indicating the time they emerged. With the increasing demands of various applications in temporal networks, it is imperative to design a general temporal network link prediction method to effectively capture the temporal characteristics of network evolution. Several temporal link prediction methods have attempted to couple spatial and temporal information. LIST\cite{LIST} characterized the network dynamics as a function of time, which used matrix factorization technique and integrated the spatial topology of network at each timestamp. By extracting target link features,  SSF\cite{SSF} used an exponential function to specify the influence of historical edges, and then combined the network structure to acquire the predictions. 
However, temporal link prediction methods based on exponential decay ignore the life cycle of information that newly added edges in the network will remain active for a certain period of time, after which the link information decays to a stable state. Besides, many real-world networks are sparse and a majority number of existing structure-based similarity methods are common neighbor related\cite{CN,CAR,CCLP,JI,RA}, 
which might cause lower performance of these methods. In addition, due to the irregular connection characteristics of the network, each node has its unique local topology. Therefore, when considering the local structure of the target link, the high-order structure of the two endpoints should also be reflected.

To address these issues, we first utilize the characteristics of the sigmoid function to systematically modify the demerits of the exponential function\cite{Sigmoid}. We propose the adjusted sigmoid function (\textit{ASF}) to quantify temporal information based on the simplified life cycle of information. 
Then, owing to the powerful mathematical representation of simplex in algebraic topology, we come up with hidden node set and latent matrix sequence, which solve the dilemma that some node pairs do not have common neighbors due to network sparsity.     
Finally, considering the endpoints asymmetry with simplex structure, it fully represents the surrounding topology information around the target links. Combining them to achieve the consistency of temporal and structural information, thus, the link prediction model TLPSS is proposed for general dynamic networks. The main contributions of this paper are as follows:

\begin{itemize}
	\item Based on the active, decay and stable states of information, we proposed a new time decay mode \textit{ASF}, which adequately considers the decay time and rate for different network information.
	
	\item We define the latent matrix sequence composed of simplex high-order structure to reveal the spatial topology of the network. The richer high-order topological information in latent edges alleviates the problem that traditional similarity methods are affected by lack of common neighbors due to the sparsity of the network.
	
	\item Coupling temporal and structural information, we introduce a temporal link prediction metric TLPSS induced by the hidden node asymmetry, and it is consistently feasible for various dynamic networks.
	
	\item We evaluate the TLPSS model on six real-world datasets and show that it outperforms other baseline models, which demonstrates the effectiveness of our proposed method.
\end{itemize}

\section*{Problem Description}
A dynamic network is defined as a graph $G_t = (V^t,E^t)$, where $V^t$ is the node set, $E^t$ is the set of links. A temporal link is denoted by $e^t(u,v)$, which means that the node $u$ and $v$ are connected at time $t\in \{1, 2, ..., T\}$.  Since this paper focuses on the link prediction, we only consider the change of edge connection with time, and fix the node set at different time as $V$. Note that node pairs are allowed to have multiple edges generated at different timestamps, and only undirected networks are concerned in this paper.

\begin{figure}[hp]
	\centering
	\includegraphics[width=15 cm]{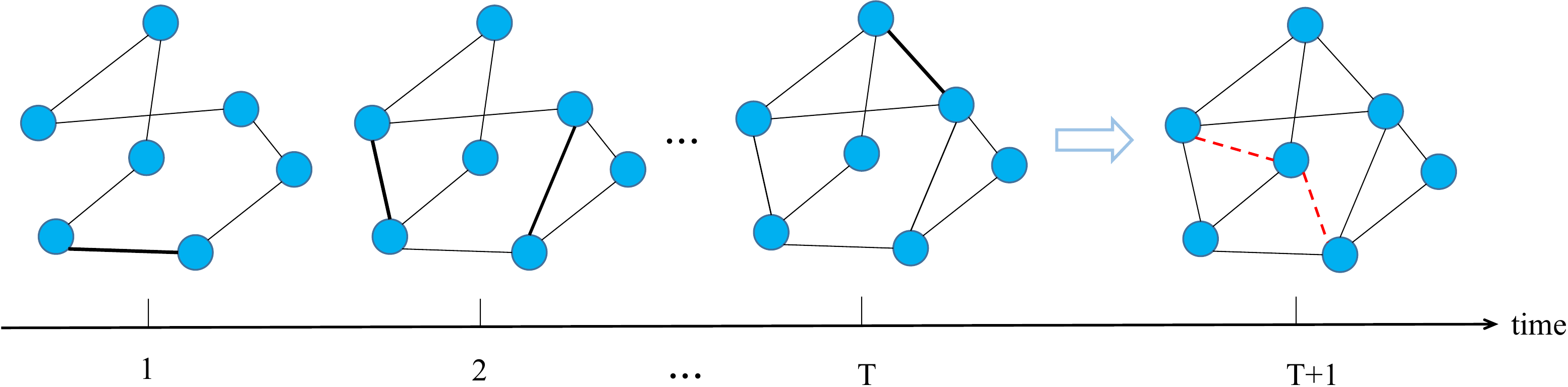}
	\caption{Schematic diagram of temporal link prediction.}
	\label{diagram}
\end{figure}

For temporal link prediction, a temporal network $G_t$ can be divided into a series of snapshots $G_t = \{G^1, G^2...,G^T\}$ at discrete time frames $\{1, ..., T\}$. For $t,s\in T$, $t < s$, $G^t$ can be regarded as the historical information of $G^s$, and they are strongly correlated and involved with the same evolution mechanism. When a set of network snapshots are given within the time period $[1, T]$,  the temporal link prediction method aims to study a function $f(\cdot)$ to predict a group of edge set $E^{T+1} = \{e^t(u,v)| u,v \in V, t = T+1 \}$ created at time $T+1$. The problem is illustrated in Fig. \ref{diagram}. Main notations in this paper are introduced in Tab. \ref{notations} for future reference. 

\begin{table}[h]
	\centering
	\begin{tabular}{|c|c|}
		\hline
		\textbf{Notation} & \textbf{Dscription}                                                                              \\ \hline
		\textbf{$G^t$}  & graph snapshot at timestamp $t$                                                                  \\ \hline
		\textbf{$A^t$}  & adjacency matrix at timestamp $t$                                                                \\ \hline
		\textbf{$B^t$}  & latent matrix at timestamp $t$                                                                   \\ \hline
		\textbf{$D^t$}  & degree matrix at timestamp 
		$t$ 															\\  \hline
		$V$             & node set                                                                                         \\ \hline
		$E^t$           & edge set at timestamp $t$                                                                        \\ \hline
		$(x\sim h)$     & latent edge, $x$ and $h$ are nodes                                                                                    \\ \hline
		$p,q$           & hyperparameters in adjusted sigmoid function                                                           \\ \hline
	\end{tabular}
	\caption{Main notations.}
	\label{notations}
\end{table}

\section*{Literature Review}
A large number of static link prediction methods have been proposed, and these methods can be divided into three categories\cite{lp_three_category}. The first category is the method based on probability statistics\cite{probability}. The basic idea of these methods is to build a parametric probabilistic model and use optimization strategies such as maximum likelihood estimation to find the optimal parameters. This type of model usually acquires satisfying results by assuming that the network has known structures or obeys specific distributions. But usually the great computational cost makes it not suitable for large-scale network, and representative works are like \cite{probability-lp1,probability-lp2,probability-lp3}. The second category is machine learning-based methods. The link prediction problem in the network can be regarded as a classification problem in machine learning\cite{machine_learning}, and the related methods work on massive training data to achieve high prediction accuracy in large-scale networks though explainable features are difficult to be extracted. Furthermore, inspired by the superiority of deep learning and graph representation learning in capturing node feature representations\cite{graph_represent}, the link prediction task can be transformed into computing distances between nodes to reveal the underlying correlation. The advantage of this type of method is that with the iterative update of representation learning algorithms, such as  deepwalk\cite{deepwalk}, node2vec\cite{node2vec} and their derivatives, the link prediction accuracy can be gradually improved, but the prediction mechanism is difficult to explain in an explicit way. The third category is the similarity-based method\cite{similarity}, which is based on the assumption that the connection probability of nodes is positively correlated with the similarity between them\cite{similarity_assumption, assumption2}. Such methods assign a score to each pair of nodes by defining a similarity function, and higher scored node pairs will have more potential to be linked together.

Recently, more complicated metrics based on temporal and structural information have been proposed for link prediction. Yu et al.\cite{LIST} proposed a link prediction model LIST with spatio-temporal consistency rule, which described the dynamic characteristics of the network as a function of time, and integrated the spatial topology of the network at each time point and the temporal characteristics of network evolution. Chen et al.\cite{STEP} proposed the temporal link prediction model STEP, which integrated structural and temporal information, and transformed the link prediction problem into a regularized optimization problem by constructing a sequence of high-order matrices to capture the implicit relationship of node pairs. Li et al.\cite{SSF} proposed a structure subgraph feature model SSF that is based on link feature extraction. This method effectively represented the topological features around the target link and normalized the influence of multiple links and different timestamps in structural subgraphs. 

Complex networks have become the dominant paradigm for dynamic modeling of interacting systems. However, networks are inherently limited to describe the interactions of node pairs, while real-world complex systems are often characterized by high-order structures. Furthermore, interaction behaviors based on structure take place among more than two nodes at once \cite{PNAS-LP}. Empirical studies reveal that high-order interactions are ubiquitous in many complex systems, and such community behaviors play key roles in physiological, biological, and neurological field. In addition, high-order interactions can greatly affect the dynamics of networked systems, ranging from diffusion\cite{diffusion}, synchronization\cite{synchronization}, to social evolution processes\cite{social_evolutionary}, and may lead to the emergence of explosive transitions between node states. For a deeper understanding of the network pattern structure, we can model it via set systems from the perspective of algebraic topology. For example, high-order structures, such as hypergraphs and simplicial complex are better tools for characterizing many social, biological systems\cite{hyper_and_Simplex}. In addition to recognize the high-order structure in the network, it is important to measure the interaction information of the different structures. Applying simplex structure to complex networks due to its powerful mathematical representation and fully quantifying structure interaction information is the key to the performance of link prediction.

\section*{Methods}
\subsection*{Time Decay Function}
The most crucial part of the temporal network link prediction task is to effectively process historical information. Based on the accumulation of historical time data, network evolution pays attention to the overall changes of the network, and performs the complex behavior of dynamic networks. Similarly, the purpose of link prediction is to understand how these characteristics will evolve over time. Link prediction makes use of temporal information to reveal the relationship between the current state of the network and its most recent states. The basic principle of dynamic link prediction is temporal smoothing\cite{time_smoothing}, which assumes that the current state of a network should not change dramatically from its most recent history in general. Several researchers concern the exponential function as time decay function\cite{SSF,LIST},
\begin{equation}
	f(s,t) = e^{-\theta (t - s)},
\end{equation}
which means the remaining influence of a history link $l$ with timestamp $s$ at present time $t$, and $\theta \in(0,1)$ is a damping factor to control the speed of decay, and as a parameter, $\theta$ needs to be pre-learned.

\begin{figure}[h]
	\centering
	\includegraphics[width=12cm]{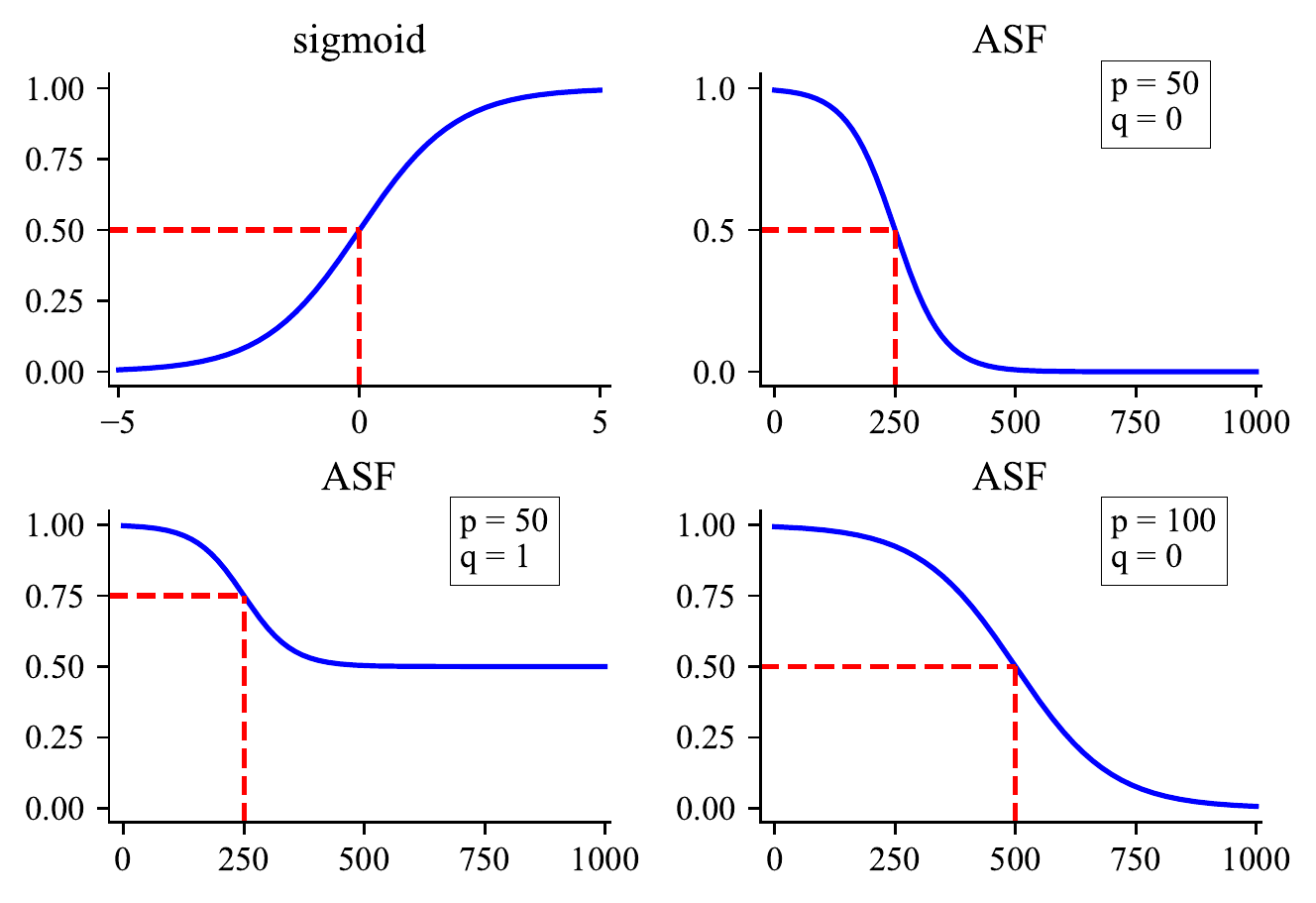}
	\caption{Sigmoid and ASF function. The upper left figure shows the original sigmoid function. The comparison of the upper right and lower left figures shows that the more information is remained with the larger parameter $q$. The comparison of the upper right and lower right figures shows that the active period of link information is determined by the parameter $p$.
	}
	\label{ASF}
\end{figure}

Choosing the exponential function as the time decay function has improved some link prediction algorithms\cite{SSF,LIST}, and it can be regarded as one of the information decay modes. Besides, scholars have been discussing the society as an information- and knowledge-based society. By giving an insight into them, information resources can be clarified with life cycle model\cite{life_cycle_1,life_cycle_2,life_cycle_3}. The life cycle phases consist of generation, institutionalization, maintenance, enhancement, and distribution. Inspired by this theory, we assume that the generation of new edges tends to remain active for a certain period of time, and then decay to a stable state. For example,
the 2022 Grammys song of the year \textit{Leave the door open}, its Billboard chart history follows the above hypothesis that it remained popular (top five) for 14 weeks, then it was gradually losing its position in the following 25 weeks, finally it fell off the chart in week number 40. Based on this assumption, we find that the sigmoid function in neural network\cite{Sigmoid} is accompanied with such properties. By parameterizing the sigmoid function, we obtain the adjusted sigmoid function (\textit{ASF}), which satisfies the assumption, as the temporal information decay function. It can be divided into active state, decay state and stable state. The formal definition of \textit{ASF} is as follows.

\begin{equation}
	ASF(x) = \frac{\frac{1}{1 + exp\{x/p - a\}} + q}{q+1},
\end{equation}
in which the parameter $p$ represents the active period of the information, and an increase in $p$ means that the information is active for a long time. For parameter $q$, it controls the decay range of information, a larger $q$ means that the lower bound of link information gets greater. 
As shown in  Fig. \ref{ASF}, the influence of the parameter $p$ is mainly reflected in the first stage of the \textit{ASF} function. Compared to the upper right figure, the lower right one has a longer active time with the larger $p$. Besides, the role of the parameter $q$ is reflected in the value range of the \textit{ASF} function. Comparison with the upper right and lower left figures indicates that the more information is remained with the larger $q$ in stable state. Unlike the former parameters, the role of parameter $a$ is only to fix the position of \textit{ASF}. In the experiment, we set $a = 5$. It is obvious that the lower bound of \textit{ASF} is $q/(q+1)$, which means that the remaining temporal information of all links in the network is greater than this value. The sigmoid fuction and variation of \textit{ASF} with parameters are illustrated in Fig. \ref{ASF}.

\subsection*{Simplex Structure in Link Prediction}
\begin{figure}[ht]
	\centering
	\includegraphics[width=12 cm]{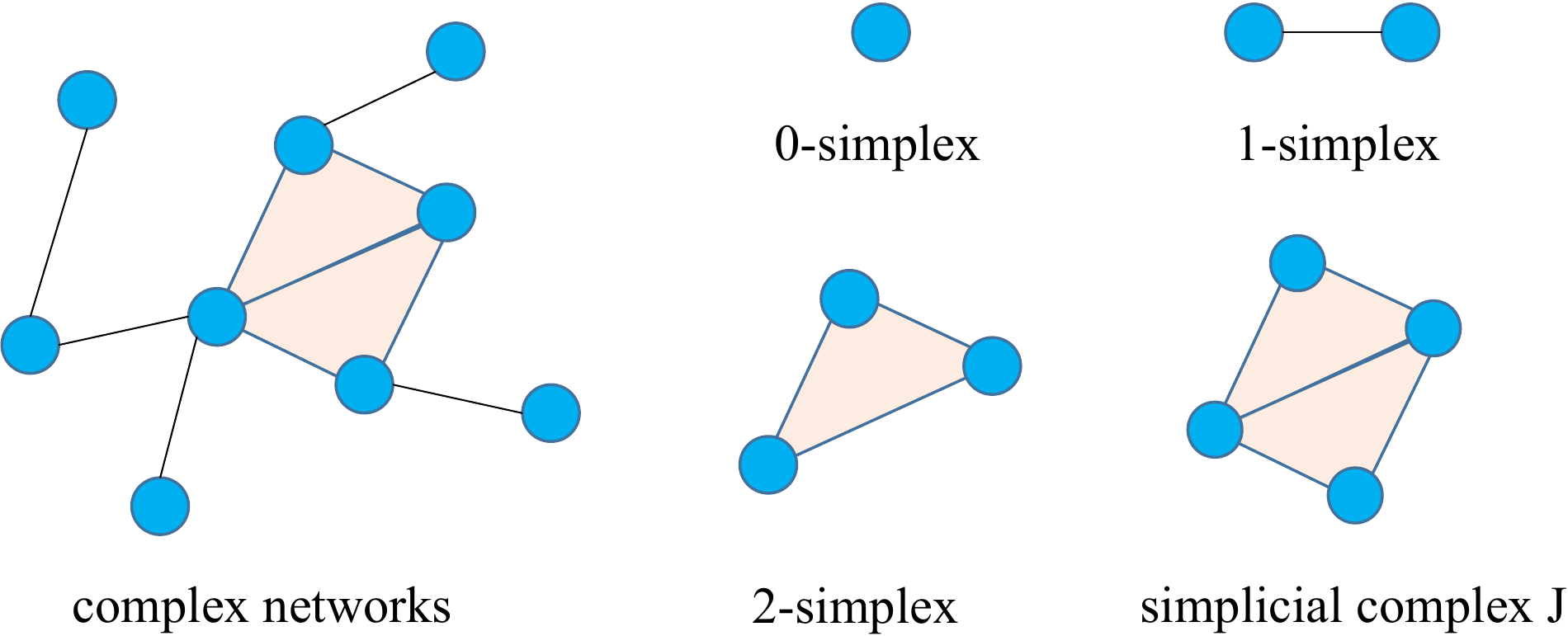}
	\caption{Examples of different types of simplex structures in networks.}
	\label{simplex}
\end{figure}

The basic premise of network model is to represent the elements of system as nodes and to use links to capture the pairwise relationships. High-order interactions mean that there are more than two entities involved at once\cite{PNAS-LP}, which are ubiquitous in various scenes\cite{h-o_scene1,h-o_scene2,h-o_scene3}. Capturing and characterizing high-order structures in networks is helpful for revealing network evolution mechanisms. Motivated by the significance of the triangular structures in network clustering and the theory of triadic closure in social networks\cite{triadic_closure}, we employ this theory via the increasing of structural order. Similar to the definition in algebraic topology, a set of $k$ nodes is called a $(k-1)$-simplex, and the set with all possible connections is called a $k$-clique in graph theory. Likewise, a simplicial complex is a finite set of simplices\cite{hyper_and_Simplex, PNAS-LP}.  As shown in Fig. \ref{simplex}, 0-simplices are nodes, 1-simplices are edges, and 2-simplices are triangles. The simplicial complex $J$ is composed of two 2-simplices.

Researchers apply high-order structure to link prediction to capture the topology information around the target link. For example, similarity metrics CAR\cite{CAR} and CCLP\cite{CCLP} use the triangle structure, which gave some insights into the mechanism of high-order interaction. But such methods mainly focus on the quantity information of triangles, while ignoring the interaction information between different structures. Simplex has been used in the study of complex dynamical systems due to its powerful mathematical representation\cite{hyper_and_Simplex}. 

In this paper, we introduce the concept of latent edge, its detailed definition is given in the following subsection. Thus, the 2-simplices around the target link form a simplicial complex structure. We measure the interaction information of these two 2-simplices to capture the local topology structure around the target link.

\subsection*{Proposed Algorithm}
Since dynamic network $G_t$ consists of a series of snapshots, adjacency matrix sequence can be expressed as $A_t =\{A^1, A^2...,A^T\}$, $A^t = \left[a^t_{i,j}\right]_{N\times N}$, $a^t_{i,j}\in [0,1]$, $N = |V|$ is the total number of nodes. Given time $t$, $a^t_{i,j}\neq 0$ means that node $i$ and node $j$ are connected, and the value is the quantification of corresponding time information by \textit{ASF} fuction. The smaller the value is, the earlier the edge is generated, and $a^t_{i,j}= 0$ means that node $i$ and node $j$ are unconnected.

In general networks, the degree of a node is defined as the number of its neighbor nodes. In our study, the elements in the adjacency matrix are no longer just 0 or 1, so the degree of a node is no longer an integer but a continuous number. Adjacency matrix can be regarded as a weighted matrix, and the weighted adjacency matrix is different at each moment. Therefore, based on the adjacency matrix sequence, the node degree information should also vary over time. The formal definition of degree matrix sequence (\textit{DMS}) is as follows.

\textbf{Definition 1.} (Degree Matrix Sequence) Given the adjacency matrix sequence, the degree information of nodes changes as the network evolves, and it can be calculated from the adjacency matrix at the corresponding snapshots. We can define \textit{DMS} as $D_t = \{D^1, D^2, ..., D^T \}$, and each degree matrix is obtained by the following calculation formula.
\begin{equation}
	D^t = [w^t(v)]_{N\times 1},  w^t(v) = \sum\limits_{z \in \Gamma(v)}A^t(v,z),
\end{equation}
in which $\Gamma(v)$ is the set of neighbors of node $v \in V$. 

The core of link prediction is correlation analysis, which reveals the intrinsic similarity between objects. A higher score of the link prediction metric indicates a higher probability of forming a link. Methods based on node centrality or common neighbors and their relevant variants indeed achieved good results\cite{CI,CN,JI}. For example, Resource Allocation index (RA) \cite{RA} considers that each node in the network has certain resources and distributes the resources equally to its neighbors. Besides, RA index shows good performance with low time complexity and high accuracy on some datasets. The formula of RA index is as follows.
\begin{equation}
	RA(x,y) = \sum\limits_{z \in \Gamma(x) \cap \Gamma(y)}\frac{1}{|\Gamma(z)|}.
\end{equation}

\begin{figure}[ht]
	\centering
	\includegraphics[width=10cm]{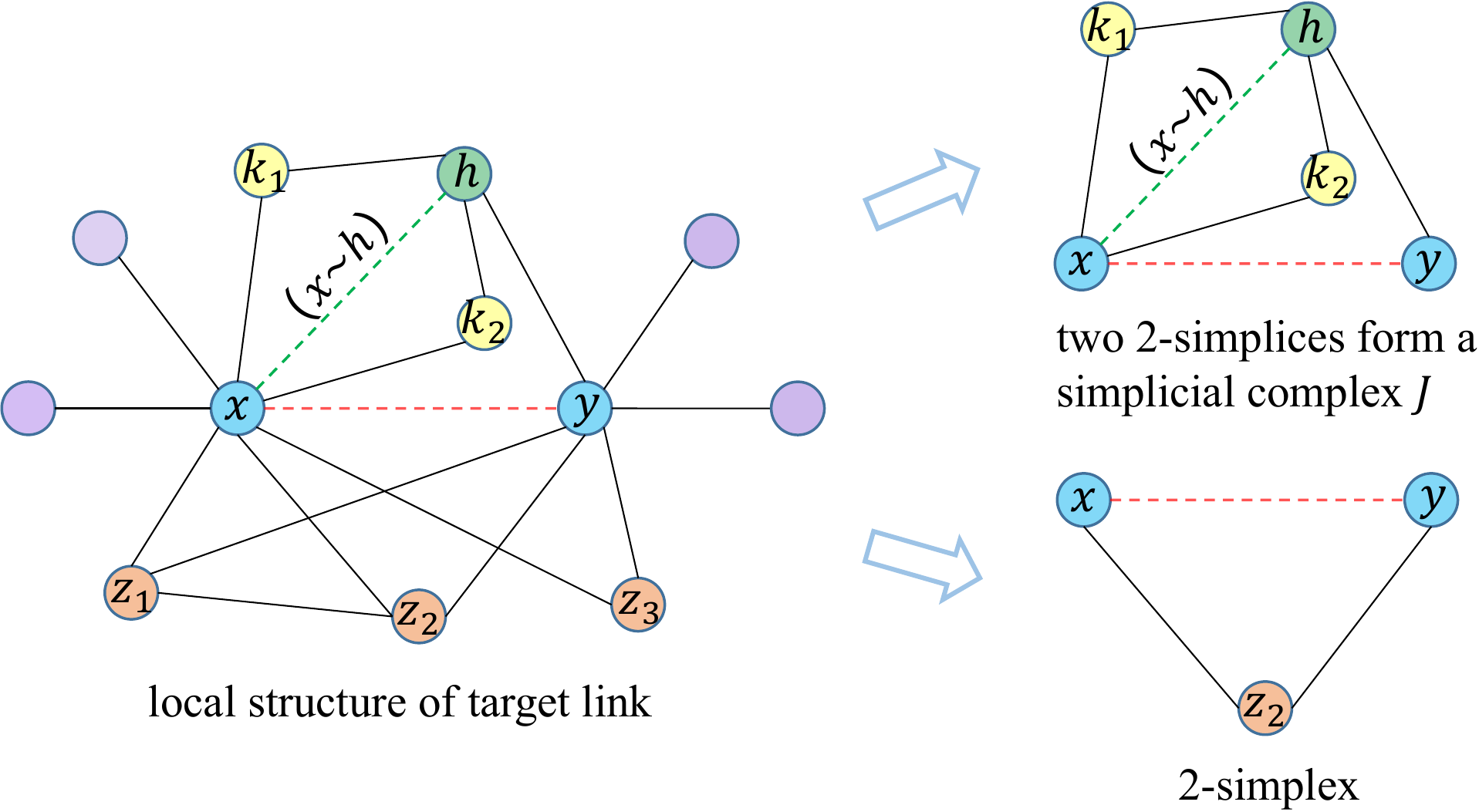}
	\caption{Schematic diagram of the topology around the target link. In this figure, the node pair $x$ and $y$ is to be predicted, $\{z_1, z_2, z_3\}$ are their common neighbors, and they form 2-simplices $\{x, z_i, y\}, i=1,2,3$. $h$ is the hidden node of endpoint $x$, and $(x\sim h)$ is the latent edge.  The simplicial complex is composed by two 2-simplices $\{x, k_1, h\}$ and $\{x, h, y\}$. Symmetrically, $k_1$ and $k_2$ are hidden nodes of node $y$.}
	\label{hidden nodes}
\end{figure}

However, this method only considers the transmission of resources through common neighbor paths, while ignoring the potential resources transmitted through local paths between two endpoints. 
As shown in Fig. \ref{hidden nodes}, for example, RA index only uses the 2-simplices $\{x, z_i, y\}, i = 1, 2, 3$, which ignores the importance of neighbor nodes of $y$ that are not directly connected to $x$. However, theses nodes participate in forming the high-order structure $J = \{x, k_i, h, y\}, i = 1, 2$ around the target link. It is assumed that the resources of node are allocated to its neighbors according to the importance of the nodes. 
The role of common neighbors in information transmission is important. But we should also pay attention to those nodes that are only directly connected to one endpoint of the target link, such as node $h$. Combining the above analysis, we define hidden node set (\textit{HNS}) for endpoints that is crucial in information transmission.\\

\textbf{Definition 2.} (Hidden Node Set) For each endpoint of target links, we define its hidden node as the kind of node that is connected to one endpoint and the neighbor of the other endpoint. Given a node pair $x$ and $y$, the \textit{HNS} of endpoint $x$ can be formulated as follows,
\begin{equation}
	H_x = \{h | h \in \Gamma(y), h\notin \Gamma(x), \Gamma(x) \cap \Gamma(h)\neq \varnothing \}.
\end{equation}
Vice versa, for endpoint $y$,
\begin{equation}
	H_y = \{h | h \in \Gamma(x), h\notin \Gamma(y), \Gamma(y) \cap \Gamma(h)\neq \varnothing \}.
\end{equation}

Based on the definition of \textit{HNS}, we can divide the neighbors of an endpoint into three categories according to their topological significance. The first is the common neighbors, the second is the hidden nodes, and the third is the rest nodes. The consideration of hidden nodes makes the link prediction method take higher-order structure into account than traditional common neighbor based similarity methods.

With the help of hidden node, we assume that there is a high probability that the hidden node will be connected to the endpoint. Therefore, we call this edge in the network that is temporarily unconnected but carries target link information as a latent edge, obviously, it is composed of endpoint and hidden node. Hidden node and latent edge play an important role in improving link prediction performance because they participate in forming a simplicial complex structure around the target link. As shown in Fig. \ref{hidden nodes}, simplicial complex structure $J$ is composed of two 2-simplices $\{x, k_1, h\}$ and $\{x, h, y\}$. Besides, their intersection is the latent edge, which contains certain information of the endpoints. We give the formal definition of latent edge (\textit{LE}) as follows.

\textbf{Definition 3.} (Latent Edge) The latent edge represents the intersection of two 2-simplices in a simplicial complex structure of the target link. As shown in Fig. \ref{hidden nodes}, the latent edge is denoted by $(x\sim h)$, in which $x$ is the endpoint and $h$ is the hidden node. 

Moreover, \textit{LE} effectively reveals the spatial topology information of the target link. It can be seen that the latent edges contain non-negligible topological information of the target node pair, but the quantification of its significance remains unsolved. Here we define the quantification strategy of such edges by the definition of latent matrix sequence (\textit{LMS}) as follows.

\textbf{Definition 4.} (Latent Matrix Sequence) The connection state of network at each snapshot can be represented by the adjacency matrix sequence $A_t =\{A^1, A^2...,A^T\}$. We define Latent Matrix Sequence as $B_t =\{B^1, B^2...,B^T\}$, the elements in the $B^t$ are latent edges. Latent edges use simplicial complex structure to fully consider the information transmission between endpoints. The value of latent edges is calculated as follows.
\begin{gather}
	B^t(i,j) = inf\{ASF\} \cdot scale\ factor \label{latent matrix},\\
	inf\{ASF\} = \frac{q}{q+1} \label{inf},\\
	scale\ factor = \frac{1}{min\{d(i), d(j)\}}\ \cdot \sum\limits_{z \in \Gamma(i) \cap \Gamma(j)}\frac{A^t(i,z)+A^t(z,j)}{(m(i,z)+m(z,j))}, \label{scale factor}
\end{gather}
where $m(i,z)$ is the number of multiple edges between node $i$ and node $z$ created at different timestampts,  $a^t_{i,j} = 0$, and $d(i)$ is the degree of a node in the traditional sense. 

These operations hold that the weight of the latent edge is less than the weight of the existing edge in the network. We simply prove it as follows.
\begin{proof}
	\vspace{-0.36cm}
	The first item $inf\{ASF\}$ in Eq. \ref{latent matrix} is the lower bound of the \textit{ASF} function, and it quantifies the time information of the connected edges in the network. 
	It is clear that the numerator of the second term in Eq. \ref{scale factor} is less than the denominator. Since $|\Gamma(i) \cap \Gamma(j)| \leqslant min\{d(i), d(j)\}$, we obtain $scale\ factor \leqslant 1$. The product of the two terms in Eq. \ref{latent matrix} ensures that the weight of the latent edge is less than the $inf\{ASF\}$. Thus, the weight of the latent edge is less than the weight of the existing edge in the network.
	\vspace{-0.40cm}
\end{proof}
The \textit{LMS} further characterizes the topology information around the target link by using simpicial complex structures. Besides, considering the complete difference between two endpoints' hidden nodes, it is important to introduce endpoints asymmetric topology information into the mechanism of link prediction. 
After the above analysis, based on the network history information from time $1$ to $T$, we can predict the generation of new links at time $T+1$. By mixing the 2-simplices information in adjacency matrix and latent matrix of endpoints, we obtain endpoints similarity scores respectively. For endpoint $x$,

\begin{equation}
	\label{s_xy}
	score(x\rightarrow y)=\sum\limits_{z_i \in \Gamma(x) \cap \Gamma(y)}\frac{A^T(x,z_i)}{w^T(z_i)} + \sum\limits_{z_i \in H_x}\frac{B^T(x,z_i)}{w^T(z_i)}.
\end{equation}
Similarly, for endpoints $y$,
\begin{equation}
	\label{s_yx}
	score(y\rightarrow x)=\sum\limits_{z_i \in \Gamma(x) \cap \Gamma(y)}\frac{A^T(y,z_i)}{w^T(z_i)} + \sum\limits_{z_i \in H_y}\frac{B^T(y,z_i)}{w^T(z_i)}.
\end{equation}

Finally, we obtain the temporal link prediction method TLPSS that integrates the features of 2-simplex topological structures, endpoints asymmetry and the \textit{ASF} time decay paradigm,
\begin{equation}
	TLPSS(x,y)=\frac{1}{2}(score(x\rightarrow y)+score(y\rightarrow x)).
\end{equation}	

\begin{figure}[ht]
	\centering
	\includegraphics[width=15 cm]{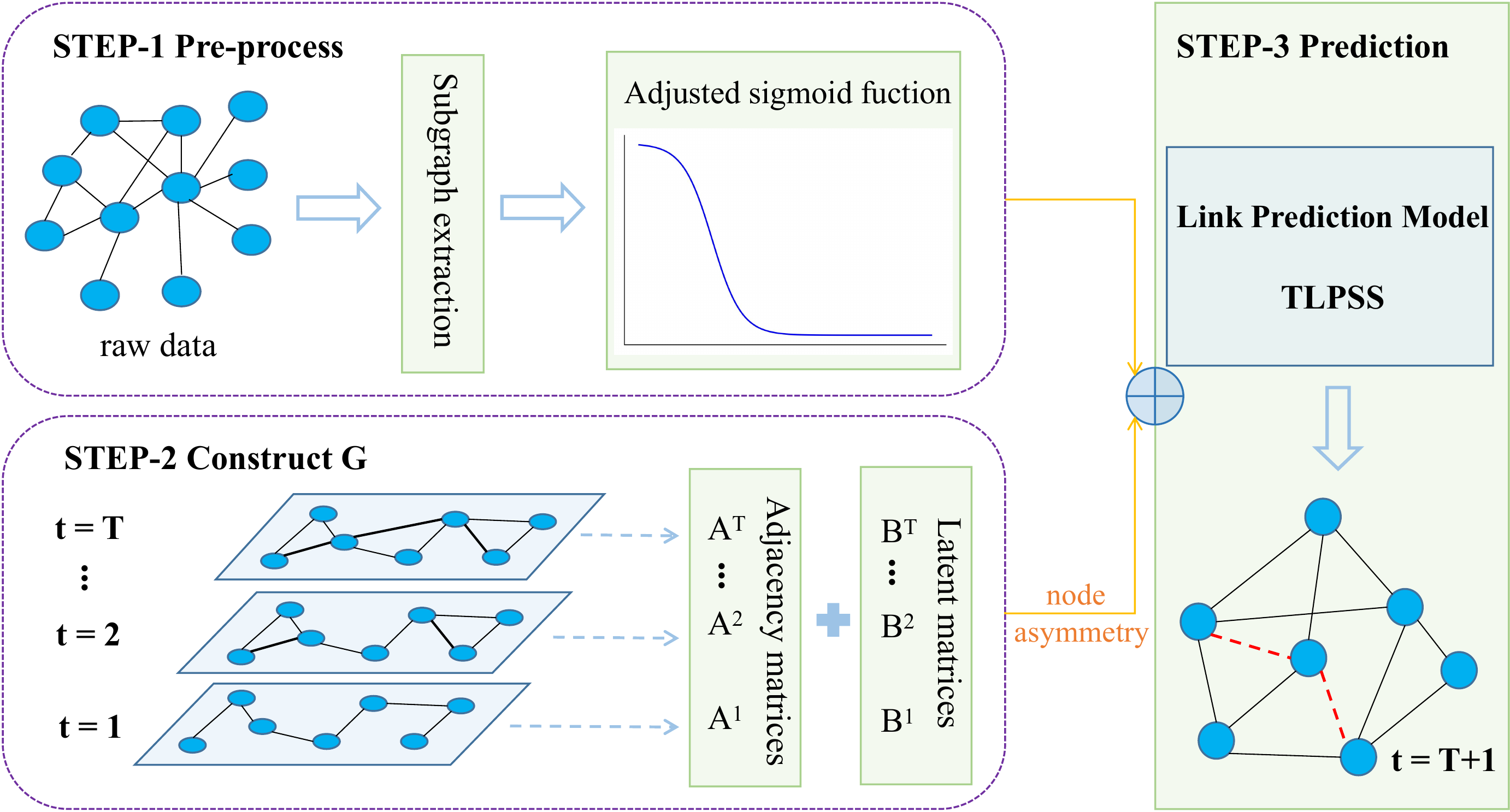}
	\caption{Diagram of the proposed model TLPSS. This model contains pre-process, construct graph and prediction steps. In first step, the data is processed and decayed by \textit{ASF}. Then, according to the network snapshots, we obtain the adjacency matrix sequence and latent matrix sequence. Finally, coupling the temporal and structural information, the temporal link prediction method TLPSS is proposed.}
	\label{model_frame}
\end{figure}

Based on the above analysis, the proposed algorithm can be divided into three steps, and the schematic diagram of TLPSS model is shown in Fig. \ref{model_frame}. Firstly, it is necessary to pre-process the data, because the data we get is always noisy. Specifically, we remove data with missing temporal information and sort them according to time evolution. The subgraph extraction strategy is used for large-scale networks to reduce computational cost. Then link information decays according to the historical time by \textit{ASF}. Secondly, by constructing the processed data, we get the weighted adjacency matrices at different timestamps and the latent matrices on this basis. Thirdly, considering node asymmetry, the data is input into our link prediction model to evaluate the generation of new links at the next time period.

\section*{Experimental Setting}
In this section, we conduct experiments to evaluate the effectiveness of the proposed approach by using six real-world datasets for link prediction tasks, and compare its performance with six baseline algorithms. At first, we briefly introduce the datasets from different domains.
\subsection*{Datasets Description}
\begin{itemize}
	\item \textbf{Contact}\cite{contact}: This network represents contacts between people, which is measured by carried wireless devices. Each node represents a person, and an edge between two persons shows that there was a contact between them.
	
	\item \textbf{DBLP}\cite{DBLP}: This is the citation network of DBLP, a database of scientific publications such as papers and books. Each node in the network is a publication, and each edge represents a citation of a publication by another publication.
	
	\item \textbf{Digg}\cite{Digg}: This is the reply network of the social news website Digg. Each node in the network is a user of the website, and each edge denotes that one user replied to another user.
	
	\item \textbf{Enron}\cite{Enron}: The Enron email network consists of emails sent between employees of Enron. Nodes in the network are individual employees and edges are individual emails.  It is possible to send an email to oneself, and thus this network contains loops.
	
	\item \textbf{Facebook}\cite{Facebook}: This network contains friendship data of Facebook users. A node represents a user and an edge represents a friendship between two users. 
	
	\item \textbf{Prosper}\cite{Prosper}: This network represents loans between members of the peer-to-peer lending network at Prosper.com. The network is directed from lender to borrower. Each edge is tagged the timestamps when the loan was occured.
\end{itemize}

All of these datasets are dynamic networks, i.e., each edge is annotated with timestamps showing the formation time. Since our main concern is whether there will be an edge between two nodes, the direction of the edge in the network is eliminated in the experiment. Tab. \ref{data info} shows major information of those datasets. \textit{Total duration} is the length of the
time span of dynamic networks, specifically, \textit{h, d, w, m} and \textit{y} stand for \textit{hour, day, week, month} and \textit{year} respectively. \textit{Snapshot number} denotes the decay times of the network divided by the time information decay period, which is determined by the edge distribution of each dataset.
Besides, we normalize the time attribute so that the timestamp of network starts from 1. In the link prediction evaluation stage, the existing link set $E_t$ in the network is divided into two sets: train set $E(T)$ and test set $E(P)$ according to time evolution. The ratio between them is around 9:1. 
\begin{table}[h]
	\begin{tabular}{|c|c|c|c|c|c|c|c|}
			\hline
			\textbf{Dataset}  & Node number & Edge number & Ave. Degree & Start date & End date   & Total duration           & Snapshot number \\ \hline
			\textbf{Contact}  & 273         & 28227       & 206.78      & 1970/1/1   & 1970/1/4   & 70h                                    & 70/h              \\ \hline
			\textbf{DBLP}     & 1169        & 10667       & 18.24        & 1986/1/1   & 1996/1/1   & 10y                                    & 10/y               \\ \hline
			\textbf{Digg}     & 3159        & 17661       & 11.18        & 2008/11/3  & 2008/11/11 & 8d                                     & 192/h             \\ \hline
			\textbf{Enron}    & 883         & 31092       & 70.42       & 2000/2/15  & 2000/6/14  & 4m                                    & 17/w              \\ \hline
			\textbf{Facebook} & 3877        & 30480       & 15.72        & 2007/11/30 & 2008/8/26  & 9m                                     & 270/d             \\ \hline
			\textbf{Prosper}  & 2561        & 46540       & 36.34       & 2006/10/10 & 2006/12/11 & 2m                                     & 60/d              \\ \hline
	\end{tabular}
	\caption{Network datasets statistics.}
	\label{data info}
\end{table}

\subsection*{Baseline Methods and Evaluation Metrics}
\textbf{Baseline Methods.} We compare our proposed model TLPSS with the following link prediction methods. These methods are usually used for static networks, it can also be applied to time-varying
networks by aggregating all edges from different timestamps to one network. We make improvements over traditional baseline methods. The number of common neighbors of target link and the number of triangular closures around them are determined by the weights of edges. The specific definitions are shown in the Tab. \ref{baseline}.

\begin{table}[h]
	\centering
	\footnotesize
	\begin{tabular}{|m{3cm}|m{7cm}|m{6cm}|}%
		\hline
		\textbf{Baseline Methods}                 &\textbf{Description}               &\textbf{Definition}                                                                         \\ \hline
		Common Neighbors (CN) & The algorithm uses the number of common
		neighbors as an indicator to measure the possibility of establishing a link between two nodes \cite{CN}.                                                                                                                                    & $CN\_ASF(x,y)=\frac{1}{2}\sum\limits_{z \in \Gamma(x) \cap \Gamma(y)}A^T_{x,z}+A^T_{y,z}$             \\ \hline
		Jaccard Index (JA)     & This algorithm evaluates the probability of connecting edges also by measuring the number of common neighbors, it is the normalized version of $CN\_ASF$ \cite{JI}.                                                                                                                                                             & $JA\_ASF(x,y)=CN\_ASF(x,y)/(w^T(x)+w^T(y))$                                                      \\ \hline
		Preferential Attachment (PA)              & In this algorithm, the probability that the target link is connected is proportional to the product of the degrees of the two endpoints, it is a hub-promoted method. \cite{PA}.                                                                                                         & $PA\_ASF(x,y)=w^T(x) \cdot w^T(y)$                                                                \\ \hline
		Resource Allocation (RA)                  & Common neighbors serve as a medium for resource transfer, and the weight of common neighbors is inversely proportional to its degree \cite{RA}.                                                                            & $RA\_ASF(x,y)=\sum\limits_{z \in \Gamma(x) \cap \Gamma(y)}\frac{1}{w^T(z)}$                       \\ \hline
		Cannistrai Alanis Ravai (CAR)             & The algorithm utilizes the links between commmon neighbors, along with commmon neighbors information, where LCL'(x,y) is total weights of links between common-neighbors \cite{CAR}.                                         & $CAR\_ASF(x,y)=CN\_ASF(x,y)\cdot LCL'(x,y)$                                                            \\ \hline
		Clustering Coefficient-based Index (CCLP) & This metric employs clustering coefficient of common neighbors to reflect the density of triangles within a local network environment, where $\Delta'$ is the total weight of weighted triangles among common-neighbors \cite{CCLP}. & $CCLP\_ASF(x,y)=\sum\limits_{z \in \Gamma(x) \cap \Gamma(y)}\frac{\Delta'}{d(z)\cdot (d(z)-1)/2}$ \\ \hline
	\end{tabular}
	\caption{Baseline link prediction methods for temporal networks.}
	\label{baseline}
\end{table}

\noindent\textbf{Evaluation Metrics.} We use two commonly adopted evaluation metrics, AUC \cite{AUC} and precision \cite{precision} to systematically evaluate the performance of the aforementioned methods. AUC can be interpreted as the probability that the similarity value of a randomly chosen new link is greater than a randomly chosen nonexistent link. A larger AUC value means better performance of the model. AUC measures the accuracy of the algorithm from a general perspective, while sometimes we  pursue how many positive items the top part of link prediction methods output contains. Precision considers whether the edge in the top-L position is accurately predicted.

\section*{Results and Discussions}
In this section, we verify the effectiveness of TLPSS in real-world datasets with different evaluation metrics. 
Tab. \ref{AUC performance} shows the AUC performance of different approaches on six dynamic networks, and the best performance on every datasets is highlighted. The proposed TLPSS model outperforms all baselines consistently across all six dynamic networks. 
The average performance of the TLPSS model outperforms other baseline methods about 15\%, especially in Digg and Prosper datasets, our model leads with 20\% and 30\% respectively. Our proposed model TLPSS can be regarded as asymmetric modification of RA if we remove the latent edge terms in Eq. \ref{s_xy} and Eq. \ref{s_yx}. Experimental results illustrate that capturing the local structure around the endpoints separately could improve the link prediction performance.
Besides, the reason for the superiority of TLPSS is that considering the latent edges in the network can address the cold-start problem for traditional common neighbor based link prediction methods, since the sparsity of the network might lead to a lack of common neighbors. In conclusion, the clear domination of the TLPSS index indicates that a deep understanding of life cycle of information and topology information could be converted to an outstanding link prediction algorithm.

\begin{table}[h]
	\centering
	\begin{tabular}{|c|c|c|c|c|c|c|c|}
		\hline
		\textbf{AUC}      & \textbf{CN} & \textbf{JA} & \textbf{PA} & \textbf{RA} & \textbf{CAR} & \textbf{CCLP} & \textbf{TLPSS}  \\ \hline
		\textbf{Contact}  & 0.9525      & 0.8611      & 0.9020       & 0.9324      & 0.9334       & 0.8495        & \textbf{0.9751} \\ \hline
		\textbf{DBLP}     & 0.8627      & 0.8591      & 0.5201      & 0.8718      & 0.7748       & 0.8105        & \textbf{0.9183} \\ \hline
		\textbf{Digg}     & 0.6426      & 0.6445      & 0.5089      & 0.6472      & 0.6525       & 0.6119        & \textbf{0.8905} \\ \hline
		\textbf{Enron}    & 0.8872      & 0.8730       & 0.4681      & 0.8852      & 0.8745       & 0.8277        & \textbf{0.9014} \\ \hline
		\textbf{Facebook} & 0.7505      & 0.7518      & 0.4731      & 0.7520       & 0.5798       & 0.7453        & \textbf{0.9093} \\ \hline
		\textbf{Prosper}   & 0.4018      & 0.4102      & 0.4639      & 0.3907      & 0.4993       & 0.4036        & \textbf{0.7493} \\ \hline
	\end{tabular}
	\caption{Comparison of the AUC value between TLPSS and baseline methods.}
	\label{AUC performance}
\end{table}

Tab. \ref{precision performance} reports the precision values of TLPSS and other similarity algorithms. Due to the different scales of datasets, for \textit{Contact, DBLP, Digg, Enron, Facebook} and \textit{Prosper}, we set $L = \{100, 100, 1000, 100, 1000, 2500\}$ respectively. It can be seen from the table that the proposed method TLPSS is superior to other methods and can provide the highest accuracy on most datasets. In Contact dataset, five of the competing methods all achieve superior performance. It shows that in densely connected network, triangular closure structures based methods like CN are already sufficient. Besides, compared with CN on Contact dataset, JA has much lower performance, this demonstrates that the normalization operation does not always work. To sum up, TLPSS model has better accuracy on most sparse networks, which indicates that the consideration of hidden node and latent edge can properly reveal the structural information around the target link.
\begin{table}[h]
	\centering
	\begin{tabular}{|c|c|c|c|c|c|c|c|}
		\hline
		\textbf{Precision} & \textbf{CN}     & \textbf{JA} & \textbf{PA} & \textbf{RA} & \textbf{CAR}    & \textbf{CCLP} & \textbf{TLPSS}  \\ \hline
		\textbf{Contact}   & \textbf{0.9900} & 0.3700      & 0.6600      & 0.9700      & \textbf{0.9900} & 0.9600        & 0.9600          \\ \hline
		\textbf{DBLP}      & 0.2100          & 0.1700      & 0.0000      & 0.0210      & 0.1400          & 0.0800        & \textbf{0.3600} \\ \hline
		\textbf{Digg}      & 0.0670          & 0.0530      & 0.0000      & 0.0240      & 0.0830          & 0.0070        & \textbf{0.0910} \\ \hline
		\textbf{Enron}     & 0.6500          & 0.6500      & 0.0000      & 0.2400      & 0.6100          & 0.1900        & \textbf{0.6800} \\ \hline
		\textbf{Facebook}  & 0.0080          & 0.0090      & 0.0000      & 0.0030      & 0.0050          & 0.0080        & \textbf{0.0100} \\ \hline
		\textbf{Prosper}    & 0.0004          & 0.0008      & 0.0024      & 0.0004      & 0.0024          & 0.0028        & \textbf{0.0032} \\ \hline
	\end{tabular}
	\caption{Comparison of the Precision value between TLPSS and baseline methods.}
	\label{precision performance}
\end{table}

\subsection*{Sensitivity Test of Parameter \textit{p} in ASF}
\begin{figure}[h]
	\centering
	\includegraphics[width=12cm]{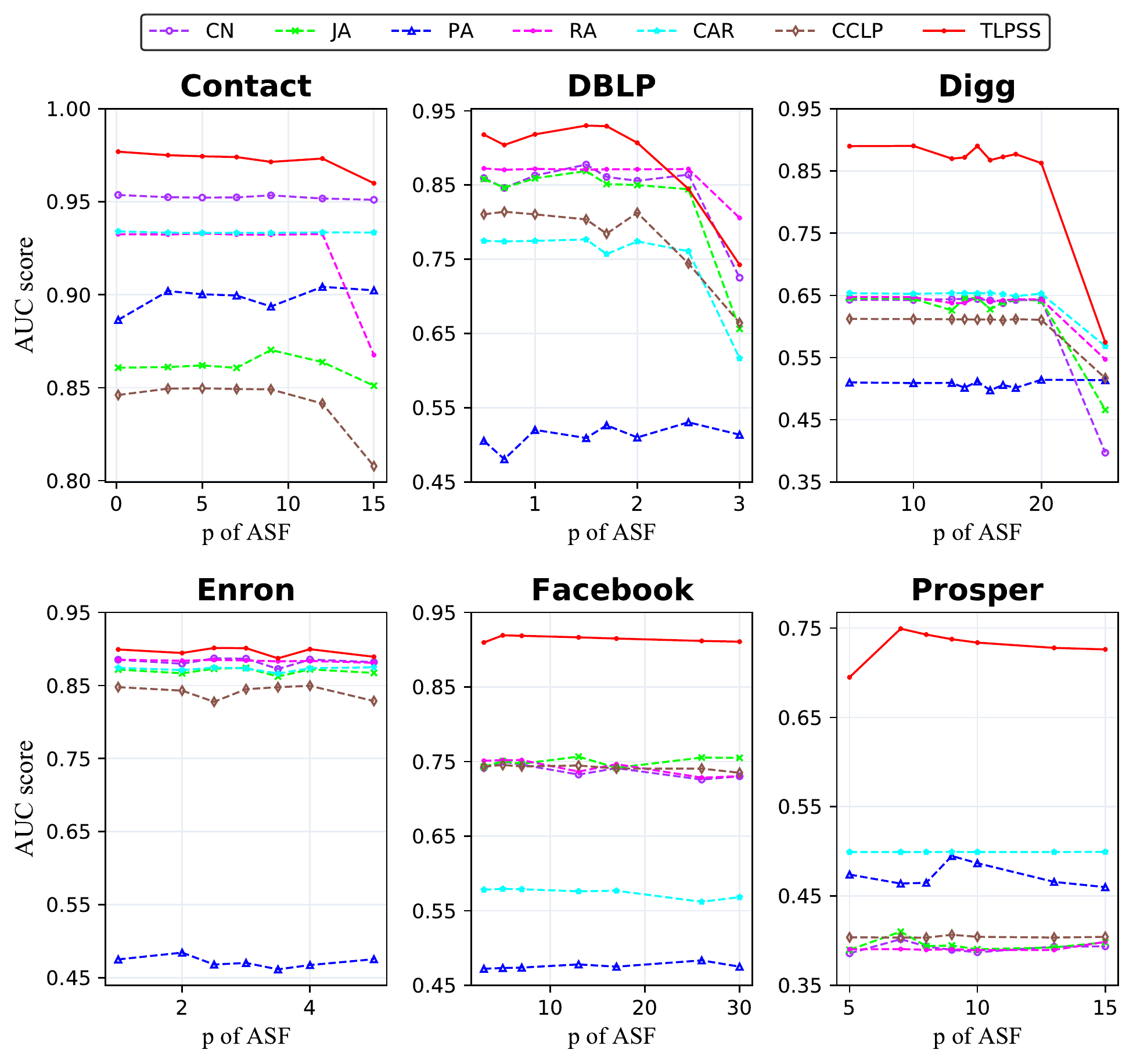}
	\caption{Performance comparison of varying parameter $p$ in different dynamic networks. All methods are based on the same temporal information decayed by \textit{ASF}. The performance of TLPSS is  superior to other baseline methods.}
	\label{comparision of p}
\end{figure}

We first study the impact of different setting of the parameter $p$ in Eq. \ref{ASF}, and set the parameter $q = 1$. Fig. \ref{comparision of p} shows the performance of different methods with varied parameter $p$ on six real-world datasets. For \textit{Contact, DBLP, Digg, Enron, Facebook} and \textit{Prosper} datasets, we obtain the optimal value of  parameter $p = \{3, 1, 10, 2.5, 5, 7\}$ respectively. There are several interesting phenomenons. First, TLPSS outperforms all other methods in most cases, and it can be interpreted that the consideration of hidden node and latent edge could unveil the spatial structure features around the target link. Second, the performance of most methods drop quickly when the parameter $p$ takes a large value on \textit{DBLP} and \textit{Digg} datasets. We hold that the large number of parameter $p$ will result in a longer decay time for the information, and inadequate utilization of temporal information due to the weight of new edge and historical edge is almost equal.Third, the optimal parameter $p$ is different for each dataset, which indicates that the decay rate of datasets in different domains can be revealed by \textit{ASF}. Forth, in \textit{Digg} and \textit{Facebook} datasets, the average degrees of these two social networks are low, common neighbor-based methods have similar performance. 
Unlike other approaches, TLPSS takes historical temporal information and simplex structure into account, thus, it has further improved the overall performance on temporal networks.

\subsection*{Performance of Latent Matrix Sequence in Real-World Networks}
\begin{figure}[h]
	\centering
	\includegraphics[width=12 cm]{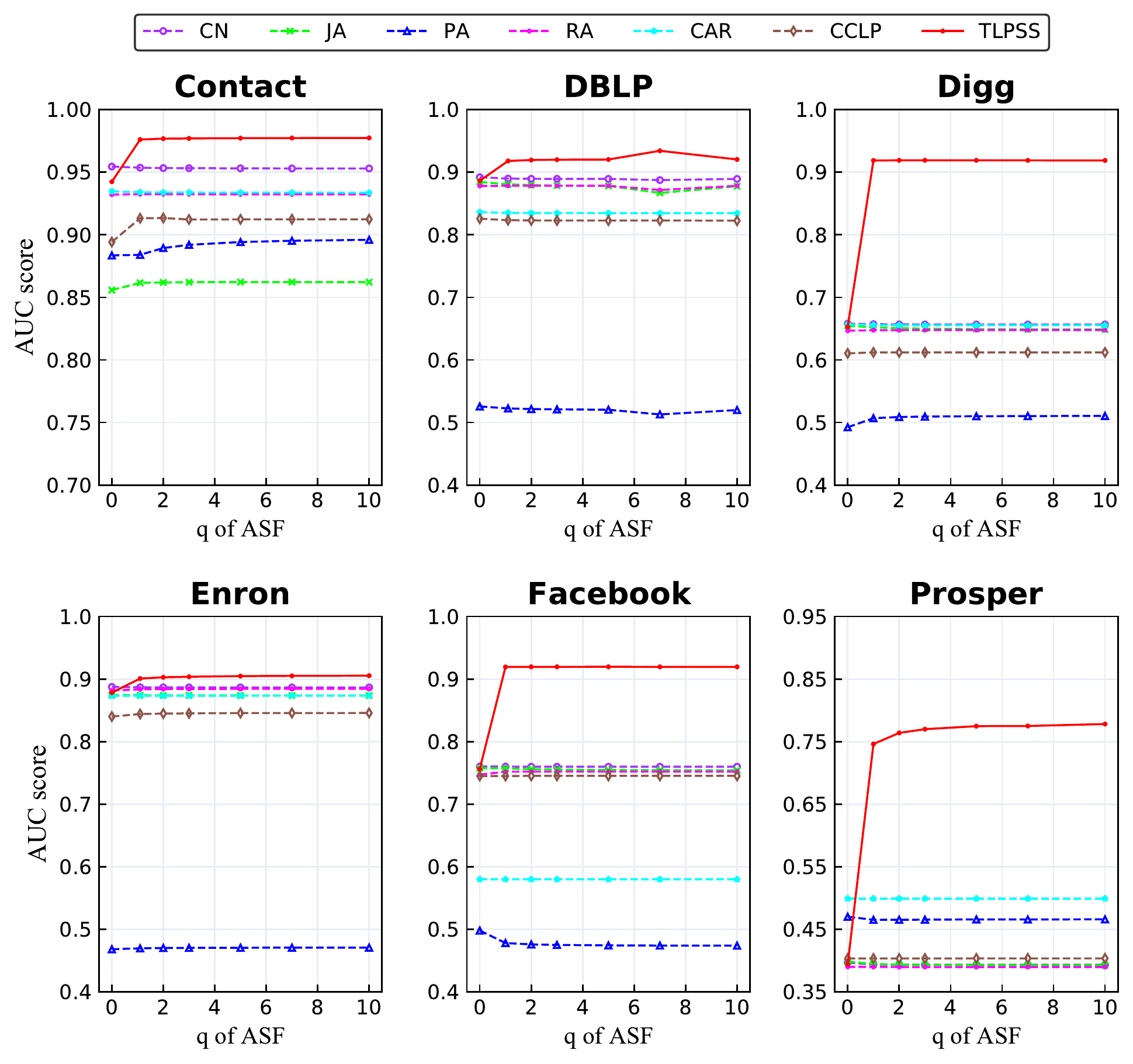}
	\caption{Performance comparison of varying parameter $q$ in different dynamic networks. The AUC value of the TLPSS model includes rising stage and stable stage. Explosive rising stage illustrates the effectiveness of latent edge composed of 2-simplices structure.}
	\label{comparision of q}
\end{figure}

Based on the definition of \textit{ASF} and \textit{LMS}, we can conclude that the weight of the latent edge in latent matrix is closely related to the parameter $q$, and its upper bound is the lower bound of \textit{ASF}, which is $q/(q + 1)$. In order to further understand the mechanism of proposed model TLPSS, the influence of parameter $q$ on AUC value is demonstrated in Fig. \ref{comparision of q}.
We set the parameter $p$ for each dataset to be the optimal value according to the former experiment. Besides, we choose different values of parameter $q$, which varies from 0 to 10 with step size 1, to compute AUC values of different algorithms. From Fig. \ref{comparision of q}, AUC value of TLPSS model fluctuates greatly when parameter $q$ increases from 0 to 1. The special case of $q=0$ indicates that there are no latent edges considered according to Eq.\ref{latent matrix}. If we remove the latent edge terms in Eq. \ref{s_xy} and Eq. \ref{s_yx}, we could see that TLPSS is the asymmetric modification of RA. This will make the 2-simplex structure composed of endpoint and hidden node to lose effect and bring damage to the performance of TLPSS method. Evidence can be found at the initial point of curves in Fig. \ref{comparision of q}, which shows that the AUC values of TLPSS and RA indexes are almost equal at a lower value.
The AUC value of the TLPSS model increases significantly when parameter $q$ varies from 0 to 1, which proves that considering latent edges in the network can address the cold-start problem. Traditional common neighbor-based link prediction methods face this problem since the sparsity of the network might lead to a lack of common neighbor.

It is clear that as $q$ increases from 0 to 10, the prediction accuracy of TLPSS increases till an optimal value, after which it maintains stable. All methods are not sensitive to the change of parameter $q$ in the stable state, this is because that the similarity scores of positive and negative samples increase proportionally, and their ordinal relationship remains unchanged. Moreover, compared with TLPSS, other similarity methods have lower average performance. It is evident that TLPSS considers the role of latent edges composed of simplex high-order structures, which makes the surrounding topological information richer.
To sum up, the large number of $q$ leads to little fluctuation on TLPSS. According to the experimental results, it is recommended to set the parameter $q = 1$.

\section*{Conclusion}
In this paper, we concentrate on the link prediction problem and design a general framework for temporal networks. We first provide a new time decay function \textit{ASF} to quantify the remaining information of different timestamps links. Next, \textit{HNS} and \textit{LE} are introduced for the target link to extract the surrounding 2-simplex high-order structures. Besides, \textit{LMS} effectively quantifies the weights of latent edges in the network, which alleviates the problem that traditional similarity methods are affected by lack of common neighbors due to the sparsity of the network. Finally, from the perspective of node asymmetry in the network, we propose the temporal link prediction method TLPSS by combining 2-simplex structural information in adjacency matrix and latent matrix. We theoretically analyze the optimality and validity of the parameters in the model. Extensive experiments on multiple datasets from different fields demonstrate the superiority of our model TLPSS compared to other baseline approaches. 
Our future work will focus on link prediction in directed temporal network with the consistency of life cycle of information and high-order structures. Also, the combination of \textit{ASF} with other types of structures extracted with deep learning methods is left for further research.

\section*{Data Availability}
All datasets in this paper are available at http://konect.cc/networks/.
\bibliography{sample}

\section*{Acknowledgements}

This work was supported by the Research and Development Program of China (Grant No.2018AAA0101100), the National Natural Science Foundation of China (Grant Nos.62141605, 62050132), the Beijing Natural Science Foundation (Grant Nos. 1192012, Z180005).

\section*{Author contributions}
R.Z. designed the research, analysed the results and wrote the paper. Q.W. prepared figures, analysed data and evaluated the algorithm. Q.Y. conducted the experiments, analysed the results and wrote the paper. W.W. designed the research, analysed the results and wrote the paper.

\section*{Competing interests}
The authors declare no competing interests.

\section*{Additional information}
\textbf{Correspondence} and requests for materials should be addressed to W.W.

\end{document}